\documentstyle[aps,prb,multicol,epsf]{revtex}

\def\Tr{\mbox{Tr}}

\newcommand{\bleq}{\ifpreprintsty
                   \else
                   \end{multicols}\vspace*{-3.5ex}{\tiny
                   \noindent\begin{tabular}[t]{c|}
                   \parbox{0.493\hsize}{~} \\ \hline \end{tabular}}
                   \fi}
\newcommand{\eleq}{\ifpreprintsty
                   \else
                   {\tiny\hspace*{\fill}\begin{tabular}[t]{|c}\hline
                    \parbox{0.49\hsize}{~} \\
                    \end{tabular}}\vspace*{-2.5ex}\begin{multicols}{2}
                    \fi}
\newcommand{\bcols}{\ifpreprintsty\else\begin{multicols}{2}\fi}
\newcommand{\ecols}{\ifpreprintsty\else\end{multicols}\fi}

 \newcommand {\be} {\begin{equation}}
\newcommand {\bea} {\begin{eqnarray} \nonumber }
\newcommand {\ee} {\end{equation}}
\newcommand {\eea} {\end{eqnarray}}
 \newcommand {\eps} {\epsilon}
 
 \newcommand {\nn} {\nonumber}
\def\(({\left(}
\def\)){\right)}
\def\[[{\left[}
\def\]]{\right]}

\begin{document}
\bibliographystyle{prsty}
\title{ Level Correlations in Disordered Metals:  the Replica $\sigma$-Model}
  
\draft

\author{Alex Kamenev$^{1,2}$ and Marc M\'ezard$^{1,3}$ 
  }
\address{$^1$Institute for Theoretical Physics, 
 University of California Santa Barbara, CA 93106-4030, USA.\\
 $^2$Department of Physics, 
 University of California Santa Barbara, CA 93106-4030, USA. \\
 $^3$CNRS, Laboratoire de Physique Th\'eorique de l'ENS,  France.
  \\
  {}~{\rm (\today)}~
  \medskip \\
  \parbox{14cm} 
    {\rm We compute   energy level correlations in weakly
    disordered metallic grains using the fermionic replica method. We
    use the standard $\sigma$--model approach  and show
    that  non--trivial saddle points, which break replica symmetry,  must be 
    included in the calculation to reproduce the oscillatory behavior of
    the correlations. We calculate  the correlation 
    functions  in all three classical ensembles GOE, GUE and GSE including 
    the finite--dimensional gradient corrections. 
    Our results coincides with those  obtained
    by the supersymmetric $\sigma$--model and the semi-classical trace formula. 
     }\bigskip \\ }

\maketitle

%\bcols

\section{Introduction}
\label{s1}

The statistics of energy levels of electrons in  disordered metals
 has attracted much attention in the past 
decades. Gor'kov and Eliashberg \cite{Gorkov64} conjectured that it obeys 
Wigner--Dyson laws derived for  random matrices \cite{Mehta}. 
This conjecture received a 
strong support
almost twenty years later when Efetov \cite{Efetov83} 
introduced the supersymmetric (SUSY)   
$\sigma$--model. It appears that the zero--dimensional  version of the 
$\sigma$--model gives exactly the random matrix theory  statistics of 
Wigner and 
Dyson. The use of the SUSY formulation seemed crucial since the alternative 
replica theory \cite{Wegner79,Efetov81}, when applied to the
pure random matrix problem, seemed unable to reproduce the correct
oscillatory behavior of the level correlation function 
\cite{Zirnbauer85}. Later 
Altshuler and Shklovskii \cite{Altshuler86} realized that in a finite 
dimensional system, the correlation function is modified with respect to 
  the universal random matrix level statistics: this modification appears when 
the 
energy difference is  of the order of the Thouless energy, $E_c$ 
(equal to $\hbar$ over the diffusion time through the sample), and the
corrections depend on the dimensionality $d$,  conductance $g$,
and  shape of the sample.
They used diagrammatic perturbation theory and consequently could trace only 
the modifications of the non-oscillatory part of the correlation functions. 
Finite dimensional modifications 
of the oscillatory part by the gradient terms  were 
calculated
in Ref. \cite{Mirlin94}, and more generally in Ref. \cite{Andreev95}, 
using the SUSY technique. The result were subsequently reproduced using 
a semi-classical 
trace formula approach \cite{Bogomolny96}. It followed from these works that 
a power--law decay of the oscillatory correlations crosses over to an 
exponential 
decay at the scale $E_c$. The precise behavior of this crossover depends both on 
the 
dimensionality and the symmetry class of problem. The essential feature of these 
results is that {\em all} non--universal terms may be expressed through the 
spectral determinant of a single classical differential operator. For the case 
of 
a disordered metal grain it turns out to be the diffusion operator in the 
corresponding geometry.   

In a recent paper \cite{KM1} we have shown  how the fermionic replica method 
may 
be used to calculate the 
level statistics of the Gaussian unitary ensemble (GUE). 
The 
calculations of Ref. \cite{KM1} were specific to the GUE and essentially 
used 
the Itzykson--Zuber \cite{Itzykson80}  integral for the unitary group. The 
purpose of the present paper is to present a more general approach to the 
fermionic 
replica calculations of the level statistics, which is not based on the 
peculiarities of the unitary ensemble, and uses rather the standard
path of the $\sigma$-model. We shall present the calculations 
of the level correlations in disordered metals for all 
three 
classical symmetry ensembles:
 orthogonal (GOE), unitary and symplectic (GSE). We also 
include the effects of  gradient terms on the level statistics reproducing 
exactly the results of Refs.~\cite{Mirlin94,Andreev95,Bogomolny96}. 
Our colleagues I.V. Yurkevich and I.V. Lerner have independently been developing
a non linear $\sigma$-model approach with replica symmetry, using 
some complementary approach to ours \cite{YuLe}.

Our strategy is as follows: we deal with 
the standard fermionic replica $\sigma$--model \cite{Efetov81} with an action 
written 
in terms of  the $(n_1+n_2)\times (n_1+n_2)$ dimensional $\hat Q$ matrix, 
where $n_{1,2} \to 0$ are numbers of replicas. 
The symmetry group of the action $G(n_1+n_2)$ is broken down to the 
exact 
$G(n_1)\times G(n_2)$ by a finite energy difference 
$\omega=\epsilon_1-\epsilon_2$ of the 
correlation function ($G$ is a symmetry--group of the $\hat Q$--matrix, which 
depends 
on a symmetry class of the problem). 
Based on the experience of the GUE 
solution \cite{KM1}, we consider  all possible saddle points of the 
$\sigma$--model both replica symmetric and replica non--symmetric. 
The latter spontaneously brake the exact symmetry of each sub-block $G(n)$ 
down to $G(p)\times G(n-p)$ with $0\leq p \leq n$ (here $n=n_1, n_2$ and  
$p=p_1, p_2$).
The corresponding manifold of the Goldstone modes has an exact degeneracy 
for space--independent (${\bf q}=0$) modes.  The contribution of 
such saddle point manifolds (the coset space $G(n)/G(p)G(n-p)$ ) 
to the partition (generating) function is proportional to their volume.
The volumes of the  coset spaces play  a central role in our analysis, since 
after the analytical continuation $n\to 0$ they determine which of the saddle 
points contribute to the generating function. It turns out that  in addition to 
the replica--symmetric (perturbative) saddle point ($p=0$) there is only one 
additional saddle point manifold (in each block) with $p=1$ in the GOE and GUE 
cases and two 
manifolds $p=1$ and $p=2$ in the GSE case. These replica--nonsymmetric saddle 
points 
give rise to an oscillatory part of the correlation function (since their action 
remains finite, and imaginary, in the $n\to 0$ limit). 
One thus gets the correct oscillatory behaviors of the
correlations, with
one oscillation frequency in the GOE and GUE and two oscillation frequencies in 
the GSE. 
%NEW
One should notice that the effect of the replica symmetry breaking saddle points
are not limited to the correlation of levels. In the random matrix limit, they
are known to describe the
exponentially small tails of  the density of
states outside the asymptotic support of the spectrum 
\cite{cav} and the finite size oscillatory correction inside \cite{KM1}. 

We then calculate the fluctuations 
 around each of the saddle point manifolds in the Gaussian approximation. This 
is 
legitimate at relatively large energy $\omega\gg \Delta$ and for a good metal,
$g\equiv E_c/\Delta \gg 1$ ($\Delta$ is the mean level spacing and 
$g$ is dimensionless conductance). No relation between 
$\omega$ and $E_c$ is assumed. As a result, one obtains the 
energy dependent amplitudes of  the oscillatory parts, as well as those of
the smooth parts, of 
the 
correlation functions, in the asymptotic regime $\omega\gg \Delta$. For small 
energy, $\omega\ll E_c$, the correlation coincides with the 
random matrix theory predictions, whereas for 
larger energy $\omega > E_c$ it get modified in the non--universal 
(dimensionality and $g$--dependent) way in agreement with  
Refs.~\cite{Mirlin94,Andreev95,Bogomolny96}. 

The structure of the paper is as follows. In section \ref{s2} we introduce 
notations 
and present a general discussion of the matrix field theory which
allows to compute the energy level correlations. We compute the saddle points 
of this action and the quadratic 
fluctuations around them.
 We also comment on the connection of our 
approach to the usual non linear $\sigma$-model. 
 In section \ref{s3}   we apply the theory to the three
classical ensembles. 
Finally in section \ref{s4} we briefly discuss the 
results, their range of validity, 
 and the possible further developments.  
The appendix contains the computations of the volumes of the relevant coset 
spaces, for each of the three symmetry classes.

\section{The replica matrix model and its  saddle points}
\label{s2}

\subsection{Preliminaries}
\label{s21}

We shall discuss the correlation functions of the density of states (DOS), which 
is defined as 
\be
\nu(\epsilon)\equiv V^{-1} \Tr\, \delta(\mu + \epsilon - H) \, ,
                                                      \label{dos}
\ee
where
$V$ is the volume and $H=H_0 + U_{dis}$ is the Hamiltonian of the system.
Here $H_0$ is the Hamiltonian of the corresponding regular (clean) system,
 and 
$U_{dis}$ is a random disorder potential.
We are interested in the large energy behavior and we thus measure 
 all energies from the large positive chemical potential 
$\mu$: the deviation $\eps$ from $\mu$ is supposed to scale as the mean level 
spacing, $\Delta$.

The retarded/advanced Green functions,  $G^{\pm}(\epsilon)$, are  defined as 
\be
G^{\pm}(\epsilon) = \big(\mu + \epsilon  - H \pm i\eta \big)^{-1}
                                                      \label{green}
\ee
with $\eta$ -- infinitesimal.  The density of states $\nu(\eps)$ is thus
equal to the small $\eta $ limit of $(G^-(\eps)-G^+(\eps))/(2\pi i)$.
The average DOS at large enough $\mu$ is a featureless smooth function, which we 
shall approximate by a constant, 
$\langle \nu(\epsilon) \rangle \equiv (\Delta V)^{-1}$. 
Hereafter the 
angular brackets stand for the averaging over the  ensemble of random 
disorder potentials, which we assume to be Gaussian and short--range correlated
with zero mean and a variance given by
\be
\langle U_{dis}(r) U_{dis}(r') \rangle = 
(2\pi\nu \tau)^{-1} \delta(r-r')\, ,
                                                      \label{dis}
\ee
where $\tau$ is an elastic scattering mean--free time. 

The main object of our study is the connected two--point correlation function 
of energy levels, defined as 
\be 
R(\epsilon_1,\epsilon_2) \equiv \nu^{-2}
\big\langle \nu(\epsilon_1) \nu(\epsilon_2)  \big\rangle -1\, .
                                                      \label{R2}
\ee
Using the fact 
that $\langle G^{\pm} G^{\pm}\rangle = 
\langle G^{\pm}\rangle\langle G^{\pm}\rangle = -(\pi V \nu)^2$,
one finds 
\be
R(\epsilon_1,\epsilon_2) =\frac{1}{2\pi^2} 
\left( \Re S(\epsilon_1,\epsilon_2) -\pi^2 \right)\,; 
\hskip .5cm
S(\epsilon_1,\epsilon_2) \equiv \Delta^2 
\big\langle G^{+}(\epsilon_1) G^{-}(\epsilon_2)  \big\rangle \, . 
                                                      \label{S2}
\ee
With the replica trick the two--point function $S$ may be written as 
\cite{Efetov81,Zirnbauer85,KM1} 
\be
S(\epsilon_1,\epsilon_2) = \lim_{n_{1,2}\to 0} 
\frac{\Delta^2}{n_1 n_2} 
\frac{\partial^2}{\partial \epsilon_1 \partial \epsilon_2} 
\langle Z^{(n_1,n_2)}(\hat E) \rangle\, ,
                                                      \label{replica}
\ee
where we have introduced the diagonal $(n_1+n_2)\times (n_1+n_2)$ matrix 
$E_{ji} = \delta_{ji} E_j$ with
\be
E_j = \left\{ \begin{array}{ll}
\mu+ \epsilon_1 + i\eta \,\,\,\,\, & j=1 , \ldots , n_1 \, ; \\ 
\mu+ \epsilon_2 - i\eta \,\,\,\,\, & j=n_1+ 1 , \ldots n_1+n_2 \, .
\end{array} 
\right.
                                                          \label{E}
\ee
The generating function $Z^{(n_1,n_2)}$ may be written as a functional 
integral over $2(n_1+n_2)$ fermionic fields. Getting from such a fermionic 
vector 
field theory
to a matrix formulation is a standard procedure \cite{Efetov81,Zirnbauer85} 
which we shall
not repeat in details. Performing the Gaussian averaging  
over $U_{dis}$, introducing a $(n_1+n_2)\times (n_1+n_2)$ 
Hubbard--Stratonovich 
matrix field $\hat Q(r)$ and integrating finally over the fermionic degrees of 
freedom,
one obtains for the generating function \cite{Efetov81,Zirnbauer85}
\be
Z^{(n_1,n_2)}(\hat E)= \int d[\hat Q] \exp \Big\{-A[\hat Q, \hat E]\Big\} \, ,
                                                           \label{Zorig}
\ee
where the action $A[\hat Q, \hat E]$ is given by 
\be
A[\hat Q, \hat E] = {\pi \nu \over 4 \tau}\,  \Tr\,  \hat Q^2
-\Tr \ln \((\hat E + {\nabla^2 \over 2 m} + {i \over 2 \tau} \hat Q\))\, .
                                                          \label{Aorig}
\ee
The symmetry of  $\hat Q$ and the integration measure $d[\hat Q]$ depend 
on the symmetry class of the  problem and will be discussed 
in section \ref{s3} separately for each ensemble.  
The trace operation includes both the replica indices and the spatial variables.

\subsection{Saddle points}
\label{s22}

  We shall evaluate the integral in Eq.~(\ref{Zorig}) by the saddle point 
method, 
and check a-posteriori that such an evaluation is indeed justified in the limit
of a weak disorder. A space independent solution of the
saddle point equation satisfies:
\be
\hat Q_{s.p.} ={i\over \pi\nu} \sum_{\bf p} \((\hat E - 
{{\bf p}^2 \over 2m} \openone +{i \over 2 \tau} \hat Q_{s.p.} \))^{-1} \ ,
                                                        \label{SPinit}
\ee 
where the sum over ${\bf p}$ runs over the set of eigenmodes of the
pure Hamiltonian, $H_0$, with some appropriate boundary conditions.
The saddle point matrix $\hat Q_{s.p.}$ may be diagonalised by a  
transformation: $\hat Q_{s.p.}= U^{-1} \hat \Lambda U$ where $U$ is an 
element of the symmetry group $G$ of the problem, $U\in G(n_1+n_2)$, 
and $\hat \Lambda$ is diagonal, $\Lambda_{ji} =\delta_{ji} \lambda_j$.
The saddle point equation (\ref{SPinit}) then implies 
that $U$ takes the form:
\be 
U=\left( 
\begin{array}{cc}
V_1  & 0 \\
0        &   V_2
\end{array} \right) \ \ \ , \ \ \ V_1 \in G(n_1)\ \ , \ \ V_2 \in G(n_2)  \ ,
                                                               \label{Usp}
\ee
while each eigenvalue is a solution of the equation:
\be
\lambda_j = {i\over \pi\nu} 
\sum_{{\bf p}} {1 \over E_j  -{{\bf p}^2 \over 2m} 
+{i \over 2 \tau} \lambda_j}\approx \mbox{sign}\ (\lambda_j)\, .
                                                              \label{lsp}
\ee
Here we have used the standard approximation, valid at large $\mu\tau $,
 where one substitutes the sum
over modes by an 
integral 
over $\epsilon_p={\bf p}^2/2m -\mu$, and  neglects  variations of
$\nu(\eps)$ in a vicinity of  $\eps=0$.
We thus find that the eigenvalues of 
$\hat Q_{s.p.}$  take the values:
\be
\lambda_j =\pm 1, \ \ \ \ j \in {1,...,n_1+n_2} \ .
                                                             \label{ev2}
\ee

There exist $(n_1+1)(n_2+1)$ saddle point manifolds. Each such
manifold ${\cal M}_{(p_1 p_2)}$ may be indexed by two integers 
$p_1 \in \{0,...,n_1\}$ and 
$p_2 \in \{0,...,n_2\}$, and is 
generated by the block-diagonal symmetry transformations  $U$ of the type 
(\ref{Usp}) applied 
to
the diagonal matrix:
\be
\hat \Lambda_{(p_1 p_2)} = \mbox{diag} \{
\underbrace{-1,\ldots -1}_{p_1}, 
\underbrace{+1,\ldots +1}_{n_1-p_1}, 
\underbrace{+1,\ldots +1}_{p_2}, 
\underbrace{-1,\ldots -1}_{n_2-p_2}  \} \, .
                                                              \label{Lambda_sp}
\end{equation} 
In order to scan the manifold ${\cal M}_{(p_1 p_2)}$ in a non redundant way, one
must restrict the symmetry transformations $U$ to
 the coset space 
\be
{\cal M}_{(p_1 p_2)} = 
\frac{G(n_1)}{G(p_1) G(n_1-p_1) } \times 
\frac{G(n_2)}{G(p_2) G(n_2-p_2) } \, .
                                                               \label{coset}
\ee
 
 It is useful to define the free propagator which is  the diagonal matrix
 with eigenvalues
\be
G_{jj} \equiv  \(( E_j + {\nabla^2 \over 2m} + 
 {i \over 2 \tau}  \lambda_j \))^{-1} \ . 
                                                            \label{freep}
\ee
The eigenvalues take four different values depending on the value of 
the 
index $j$. One can characterize them by two binary indices $(\alpha,\sigma)$, 
where 
$\alpha=1,2$ designates two replica blocks with the energies $\epsilon_{\alpha}$ 
and  $\sigma=\mbox{sign}(\lambda_j)$. 
For each energy ($\alpha=1,2$), we have a retarded and an advanced propagator, 
$G_\alpha^\pm$, defined by:
\be
G_{\alpha}^\pm \equiv \(( \mu + {\nabla^2 \over 2m} +\eps_{\alpha} \pm  
 {i \over 2 \tau}\))^{-1} \ ;\,\,\,\,  \alpha=1,2\, .
                                                             \label{freep1}
 \ee
In these notations the free propagator takes the form 
 \be
\hat G = \mbox{diag} \{
\underbrace{G_1^-,\ldots G_1^-}_{p_1}, 
\underbrace{G_1^+,\ldots G_1^+}_{n_1-p_1}, 
\underbrace{G_2^+,\ldots G_2^+}_{p_2}, 
\underbrace{G_2^-,\ldots G_2^-}_{n_2-p_2}  \} \, .
                                                             \label{freep2}
\ee

\subsection{ Saddle point action}
\label{s23}

On the manifold ${\cal M}_{(p_1 p_2)}$, the saddle point action is given by:
\be
A_{(p_1 p_2)} = {\pi \nu \over 4 \tau}\, \Tr\, \hat \Lambda_{(p_1 p_2)}^2 -
\Tr \ln \((\hat E + {\nabla^2 \over 2m} + 
{i \over 2 \tau} \hat \Lambda_{(p_1 p_2)} \)) \ ,
                                                           \label{A1}
\ee
where the trace involves both space and replica indices. In terms of the
free propagators $G_\alpha^\sigma$ defined in Eq.~(\ref{freep1}) it reads:
\be
A_{(p_1 p_2)} = {\pi(n_1+n_2) \over 4 \tau \Delta } 
     -p_1 \Tr \ln G_1^-  - (n_1-p_1) \Tr \ln G_1^+ -
      p_2 \Tr \ln G_2^+  - (n_2-p_2) \Tr \ln G_2^- \, , 
                                                            \label{A2}
\ee
where the traces involve only spatial variables. 
Expanding to  the first order in $\tau \eps_\alpha \ll 1$, and omitting 
unessential
constant factors, one finds  
\be 
\Tr \ln G_\alpha^\pm \approx 
\mp i \pi \eps_\alpha/ \Delta
                                                           \label{trln}
\ee
Finally, neglecting all constants, which vanish in the limit $n_{1,2} \to 0$, 
one obtains for the saddle point action:
\be
A_{(p_1 p_2)} = {i \pi \over \Delta } 
\(( n_1 \eps_1 -n_2 \eps_2 -2 p_1 \eps_1 + 2 p_2 \eps_2 \))
                                                           \label{Afin}
\ee

\subsection{Quadratic fluctuations}
\label{s24}

Let us expand around the saddle point 
$\hat Q_{s.p.}=U^{-1} \hat \Lambda_{(p_1 p_2)} U$, 
writing $\hat Q(r)=\hat Q_{s.p.}+U^{-1} \delta \hat Q(r) U$. 
The action expanded to the second order is
diagonalised in terms of the Fourier components $\delta Q_{ij}({\bf q})$
of the fluctuations: 
\be
A[\hat Q, \hat E] \approx A_{(p_1 p_2)} + 
{1 \over 2} \sum_{i,j=1}^{n_1+n_2} \sum_{\bf q } 
M_{ij}({\bf q}) 
\delta Q_{ij}({\bf q})  \delta Q_{ji}(-{\bf q})  
                                                       \label{quad}
\ee
where  the eigenvalue of the eigenmode $(i,j, \bf q)$, $M_{ij}({\bf q})$, is 
given by:
\be
M_{ij}({\bf q}) ={\pi\over 2 \tau\Delta}  
- {1 \over 4 \tau^2} \sum_{\bf p}
G_{ii} \left( {\bf p} + {{\bf q} \over 2} \right)  
G_{jj} \left( {\bf p} - {{\bf q} \over 2} \right) \ .
                                                      \label{mass}
\ee
There exist  a-priori sixteen different fluctuation eigenvalues
for each momentum mode ${\bf q}$.  It
is convenient to index them according to the binary decomposition introduced 
after  
Eq.~ (\ref{freep}). Each index $j \in {1,...,n+n'}$ is associated with a
pair of indices $(\alpha,\sigma)$, where $\alpha =1,2$ characterizes the energy, 
$\eps_\alpha$, and $\sigma=\mbox{sign}(\lambda_j)$ characterizes the 
retarded/advanced nature of the propagator. 
The sixteen different fluctuation eigenvalues are then:
\be
M_{(\alpha \sigma)(\alpha' \sigma')}({\bf q}) = 
{\pi  \over 2 \tau\Delta}  
- {1 \over 4 \tau^2} \sum_{\bf p}
G_\alpha^\sigma       \left( {\bf p} + {{\bf q} \over 2} \right)    
G_{\alpha'}^{\sigma'} \left( {\bf p} - {{\bf q} \over 2} \right) \ .
                                                                \label{Mev}
\ee
The corresponding momentum sums are easily computed, resulting in:
\bea 
&&\sum_{\bf p}
G_\alpha^\pm       \left( {\bf p} + {{\bf q} \over 2} \right)    
G_{\alpha'}^{\pm} \left( {\bf p} - {{\bf q} \over 2} \right) \approx 0 \,; 
                                                  \label{sums} \\
&&\sum_{\bf p}
G_\alpha^\pm       \left( {\bf p} + {{\bf q} \over 2} \right)    
G_{\alpha'}^{\mp} \left( {\bf p} - {{\bf q} \over 2} \right) \approx  
{2 \pi\tau \over \Delta}\Big[1 - 
D {\bf q}^2 \tau  \pm  i(\eps_\alpha-\eps_{\alpha'})\tau \Big] \ .
\eea
In the last expression we have expanded the sum to  first order in 
the small parameters $D {\bf q}^2 \tau \ll 1$ and 
$|\eps_1-\eps_2| \tau \ll 1$ where $D$ is the diffusion constant defined as 
$D = 2\mu \tau / (m d)$.

We obtain eventually the following list of 
eigenvalues for each spatial mode ${\bf q}$:
\begin{itemize}

\item
When $\sigma=\sigma'$ the eigenvalue is 
\be
M_{(\alpha \sigma)(\alpha' \sigma)} = {\pi  \over 2 \Delta}
\,   {1\over \tau} \ .
                                                         \label{massiv}
\ee
We shall call the corresponding modes ``massive'' and denote their 
number for each spatial mode ${\bf q}$ as ${\cal N}_m$.

\item
When $\sigma \ne \sigma'$ and $\alpha \ne \alpha'$ the eigenvalues are:
\bea
M_{(1 +)(2 -)}&=& M_{(2 -)(1 +)} ={\pi  \over 2 \Delta } 
(D {\bf q}^2 - i \omega)              \nonumber    \\ 
M_{(1 -)(2 +)}&=& M_{(2 +)(1 -)} ={\pi  \over 2 \Delta } 
(D {\bf q}^2 + i \omega)\ ,
                                                    \label{soft}
\eea
where $\omega=\eps-\eps'$. These are standard diffusive  modes 
associated 
with the $G(n_1+n_2)$ symmetry of the action, which is explicitly 
broken by a non zero $\omega$. 
We shall call them  ``soft'' modes and denote their 
number for each spatial mode ${\bf q}$ as ${\cal N}_{s-}$ and 
${\cal N}_{s+}$ correspondingly.  

\item
When $\sigma \ne \sigma'$ and $\alpha=\alpha'$ the eigenvalue is:
\be
M_{(\alpha +)(\alpha -)}= M_{(\alpha -)(\alpha +)} = 
{\pi  \over 2 \Delta } D {\bf q}^2 \ .
                                                    \label{zero}
\ee
These are the Goldstone modes associated with the spontaneous breaking of the 
exact $G(n_1)\times G(n_2)$ symmetry by  replica non--symmetric saddle points.
They exist only for the manifolds with non--zero $p_1$ or $p_2$.  
We shall call the corresponding modes ``zero'' modes and denote their 
number for each spatial mode ${\bf q}$ as ${\cal N}_z$. 

\end{itemize}

 Such a separation of modes into massive, soft and zero 
 is well justified in the limit where 
$\Delta \ll \omega;\ D {\bf q}^2 \ll 1/\tau$. 
This specifies the regime where our methods and results are applicable. 
The number of modes depends on the 
number of independent degrees of freedom of the $\hat Q$ matrix and should be 
specified separately for each of the ensembles.
We can perform now the Gaussian integrals over $\delta \hat Q$
 fluctuations. Each eigenmode with eigenvalue
$M_{(\alpha \sigma)(\alpha' \sigma')}({\bf q})$ contributes a factor 
\be 
\sqrt{\frac{\pi}{M_{(\alpha \sigma)(\alpha' \sigma')}({\bf q})} } \, 
                                                   \label{factor}
\ee
to the generating function $\langle Z^{(n_1,n_2)}\rangle $. 
The exception is the zero mode 
in the space independent, ${\bf q} = 0$, sector \cite{foot1}. 
This mode has identically 
zero mass, originating from the exact degeneracy of the ${\cal M}_{(p_1 p_2)}$ 
saddle point manifold. Therefore, in the ${\bf q} = 0$  sector the integral 
over the zero mode  results in the volume 
${\cal V}^{(p_1 p_2)}_{(n_1 n_2)}$ of the coset space (\ref{coset}). These 
volumes
are calculated in appendix for each of the three classical symmetry ensembles.

\subsection{Generating function}
\label{s25}

Finally putting all the factors together one finds for the average generating 
function 
\bea
\langle Z^{(n_1,n_2)}(\hat E)\rangle = 
&&\sum\limits_{p_1,p_2=0}^{n_1,n_2} 
e^{ -A_{(p_1,p_2)}  }
{\cal V}^{(p_1 p_2)}_{(n_1 n_2)} 
\left(\frac{2\Delta}{i\omega}\right)^{{\cal N}_{s+}/2}
\left(\frac{2\Delta}{-i\omega}\right)^{{\cal N}_{s-}/2}
(2\Delta\tau)^{{\cal N}_m/2}\times 
                                               \label{Ztot} \\
&&\prod\limits_{{\bf q}\ne 0}\left[
\left(\frac{2\Delta}{D{\bf q}^2 }\right)^{{\cal N}_z/2}
\left(\frac{2\Delta}{D{\bf q}^2+ i\omega}\right)^{{\cal N}_{s+}/2}
\left(\frac{2\Delta}{D{\bf q}^2- i\omega}\right)^{{\cal N}_{s-}/2}
(2\Delta\tau)^{{\cal N}_m/2}
\right]\, ,
%                                             \nonumber
\eea
where the saddle point action, $A_{(p_1,p_2)}$ is given by Eq.~(\ref{Afin}).
The first line in this expression represents the saddle point action and 
 the fluctuations in the ${\bf q} = 0$  sector, whereas the second line 
originates from the Gaussian fluctuations of ${\bf q} \ne 0$ modes. 
In section \ref{s3} we shall evaluate the coset space volumes,
${\cal V}^{(p_1 p_2)}_{(n_1 n_2)}$ and the
number of modes, ${\cal N}_{z,s,m}={\cal N}_{z,s,m}(p_1,p_2)$, 
for each of the classical  symmetry classes.  Hereafter we shall put $\Delta 
=1$, 
implying that all energies are measured in units of the mean level spacing, 
$\Delta$.

\subsection{The non-linear sigma model}

Let us briefly comment on  the connection to the usual formulation of the 
problem
in terms of the  non-linear $\sigma$-model. For simplicity we discuss only the 
unitary
case.
This basically amounts to a reorganization of the 
computation we did above, which uses the strong hierarchy
of masses (the massive modes are much more massive than the soft ones).
Assuming first that $1/(\tau \Delta)$ is large, one finds that the saddle points 
of Eq.~(\ref{Zorig}) are given by the set of matrices with $\hat Q^2=1$. This 
set is 
actually
an ensemble of $n_1+n_2+1$ disconnected manifolds ${\cal S}_r$, corresponding to 
all
possible  values of  $r=p_1-p_2$ (or equivalently of the trace of $\hat Q$).
It is easily seen that all the modes which move away 
from
these manifolds are massive, with a mass $\pi/(2 \Delta \tau)$. 
These massive
modes correspond to perturbing the matrix $\hat Q$ by a $\delta\hat  Q$ such 
that
$\hat Q \delta \hat Q + \delta \hat Q \hat Q \ne 0$, 
and the number of such massive modes is:
$n_1^2+n_2^2 + 2 r (r-n_1+n_2)$. 
Performing the
integration over the massive modes one can write (up to irrelevant constants):
\be
Z^{(n_1,n_2)}(\hat E) \simeq \sum_{r=-n_2}^{n_1} 
\left( \sqrt{{2 \Delta \tau \over \pi}}\right)^
{[n_1^2+n_2^2 + 2 r (r-n_1+n_2)]K}
\int_{{\cal S}_r} d[\hat Q] \exp \{ -A[\hat Q, \hat E]\} \ ,
                                                           \label{s-model}
\ee
where $K$ is the number of different ${\bf q}$ modes. 
Notice that the manifold ${\cal S}_r$ is characterized by 
$\hat Q^2=1$, $\Tr\,\hat Q= n_1-n_2- 2 r$.
It thus contains all matrices of the type 
$\hat Q =U^{-1} \hat \Lambda_{(p_1,p_2)} U$ with
$U \in G(n_1+n_2)$, and $\hat \Lambda_{(p_1,p_2)}$
defined in Eq.~(\ref{Lambda_sp}), with  $p_1-p_2=r$.

Expanding the action for slow spatial variations of $\hat Q$ on the manifold 
${\cal S}_r$,
one gets to first order in $\omega$ the standard sigma model:
\be
Z^{(n_1,n_2)}(\hat E) \simeq \sum_{r=-n_2}^{n_1}
\left( \sqrt{{2 \Delta \tau \over \pi}}\right)^
{[n_1^2+n_2^2 + 2 r (r-n_1+n_2)]K}
\int_{{\cal S}_r} d[\hat Q] \exp \Big\{  
-{\pi \nu D \over 4}{(\nabla \hat Q)^2 } - 
{i \pi \nu \omega \over 2}\, \Tr\, (\hat \Lambda \hat Q) \Big\} \ ,
                                                       \label{sigma_model_Z}
\ee
where $\hat \Lambda \equiv \hat \Lambda_{(0,0)}$.

For large $\omega$, one can study the sigma model by a saddle point 
approximation.
The generic variations around a point $\hat Q$ of ${\cal S}_r$, 
staying on ${\cal S}_r$, are of the
type $\delta \hat Q= [\hat Q,W]$, with an arbitrary matrix $W(r)$. 
The stationarity of the action 
imposes $\Tr (\hat \Lambda [\hat Q,W]) =0$, which implies that the saddle points 
$\hat Q_{s.p.}$ commute with $\hat \Lambda$ . 
One easily deduces that, on
 ${\cal S}_r$,  the saddle points 
sub-manifolds  are exactly the sub-manifolds ${\cal M}_{p_1,p_2}$
with $p_1-p_2=r$. This approach basically reorganizes our previous computation 
by grouping together all
the sub-manifolds with a fixed value of $\Tr\, \hat Q$ (or $p_1-p_2$). 
As we shall see below,  for $n_1,n_2 \to 0$, 
the only saddle points which contribute to leading
order in $\Delta \tau$ are the ones with $p_1=p_2$, which are all located
on the same manifold with $r=0$. Hence to the leading order one can
approximate the generating function in (\ref{sigma_model_Z})
by an integral over the single manifold ${\cal S}_0$, which is
what is usually done in the $\sigma$-model approach. We control this result
well at large $\omega$ because we can do the sums over the $p_1$, $p_2$ and
control the analytic continuation. But we believe that it is probably
correct also for any $\omega$. The reason is
the following: in Eq.~(\ref{sigma_model_Z}) one may extend the sum over $r$ to
a sum going from $-\infty$ to $\infty$, because when $r$ is
outside of the interval $\{ -n_2,...,n_1 \}$ the volume of the integration
space vanishes (this can be checked e.g. in the limit $\omega \to 0$). One may 
then 
take the limit $n_1,n_2 \to 0$ at fixed $r$. It is clear that the leading term
comes from $r=0$, which minimizes the number of massive modes. So the
usual $\sigma$-model formulation, with an integral over ${\cal S}_0$ only and 
the 
action given by Eq.~(\ref{sigma_model_Z}), 
seems to be correct.  However one must keep in mind that, on this manifold there
are, for large $\omega$,
several saddle point sub-manifolds, which lead to the oscillations
in the correlation functions. 
 In the random matrix case, at least, (without the gradient term)
 one may also try to perform the integration over 
entire manifold ${\cal S}_0$, without resorting to the saddle point method. If
the non linear $\sigma$-model, formulated  on  ${\cal S}_0$ only, is indeed
correct, this should give the exact result, not
restricted to $\omega \gg 1$. This procedure was attempted
 in Ref.~\cite{Zirnbauer85}, but 
  the analytical continuation of the expressions 
emerging 
from these calculations still remains to be studied.

\section{Correlation functions}
\label{s3}

\subsection{Unitary ensemble} 
\label{s31}

In the presence of a weak magnetic field the Hubbard--Stratonovich matrix $\hat 
Q$ 
is Hermitian \cite{Efetov81}. The corresponding symmetry group 
is the unitary group, $G=U$.  
The measure of the functional integral over 
hermitian matrices $Q_{ij}({\bf q})$ in Eq.~(\ref{Zorig})  
is given by 
\be
d[\hat Q] = \prod\limits_{\bf q} \[[ \prod\limits_{j} dQ_{jj}({\bf q}) 
\prod\limits_{i<j} d \Re Q_{ij}({\bf q}) d \Im Q_{ij}({\bf q}) \]] \, .
                                                  \label{u-meas}
\ee
There are $(n_1+n_2)^2$ degrees of freedom for each spatial mode, ${\bf q}$. 
Looking at the classification of modes, 
Eqs.~(\ref{massiv})--(\ref{zero}),  
one finds that the number of massive, soft and 
zero modes is 
\bea  
&&{\cal N}_m =
p_1^2+(n_1-p_1)^2+p_2^2+(n_2-p_2)^2+2 p_1 (n_2-p_2)+2 p_2 
(n_1-p_1)=n_1^2+n_2^2+2(p_1-p_2)
(p_1-p_2-n_1+n_2) \, ; \\
&&{\cal N}_{s+} = 
2 p_1 p_2  \, ;\nonumber \\
&&{\cal N}_{s-} = 
2 (n_1-p_1) (n_2-p_2) \, ;\nonumber  \\
&&{\cal N}_z = 
2 p_1(n_1-p_1) +2 p_2 (n_2-p_2) \, ,
                                                 \label{u-modes}
\eea
which add up to $(n_1+n_2)^2$ as they should.
Notice  that the number of zero modes, ${\cal N}_z$, 
coincides with the number of dimensions 
of the degenerate coset space manifold 
${U(n_1)}/({U(p_1) U(n_1-p_1) }) \times 
{U(n_2)}/({U(p_2) U(n_2-p_2) }) \, .
$
The volume of this  coset space is calculated in the appendix and is given by 
\be
{\cal V}^{(p_1 p_2)}_{(n_1 n_2)} = 
(4\pi)^{p_1(n_1-p_1)+p_2(n_2-p_2)} 
\ F_{n_1}^{p_1} \ F_{n_2}^{p_2} \, ,
                                                          \label{u-volume}
\ee
where 
\be
 F_n^p \equiv {\Gamma(1+n) \over \Gamma(1+p) \Gamma(1+n-p)} 
 \prod\limits_{j=1}^p { \Gamma(1+j)  \over \Gamma(1+(n-j+1))} \, .
                                                           \label{u-F}
\ee
Since $F_n^{p>n} = 0$ the sums over $p_1$ and $p_2$ in Eq.~(\ref{Ztot}) may be 
extended up to infinity. The resulting expression may be then continued 
analytically 
to $n_{1,2} \to 0$, using the procedure which was detailed 
in our previous paper 
\cite{KM1}.  The analytical continuation, $n\to 0$ at fixed $p$, 
of the  $F_n^{p}$ symbol, Eq.~(\ref{u-F}), shows that 
\be 
F_{n\to 0}^{0} = 1\,; \,\,\,\, F_{n\to 0}^{1} = n\,; \,\,\,\, 
F_{n\to 0}^{p\geq 2} = O(n^p)\, .
                                                          \label{u_Fcont}
\ee
Therefore only the terms with $p_{1,2} = 0,1$ may contribute to the correlation 
function $S$, Eq.~(\ref{replica}). 
The number of massive modes, in the limit 
$n_{1,2} \to 0$ at fixed $p_1,p_2$, is ${\cal N}_m\to 2 (p_1-p_2)^2$. 
Therefore
the terms with 
$p_1\ne p_2$ can be neglected to leading order  in the parameter $\Delta\tau \ll 
1$.
One thus 
ends up with the two contribution to the generating function: $p_1=p_2=0$
and $p_1=p_2=1$. 

The first piece with $p_1=p_2=0$ is the usual replica symmetric contribution.
Using Eqs.~(\ref{Ztot}), (\ref{u-modes}) and (\ref{u_Fcont}) one 
finds:
\be
\langle Z^{(n_1,n_2)}(\hat E)\rangle|_{p_1=p_2=0} = 
e^{\pi i(n_2 \eps_2-n_1 \eps_1)} 
\prod\limits_{{\bf q}} \(( {1 \over D {\bf q}^2-i \omega}\))^{n_1n_2}\, .
                                                         \label{U-Z1}
\ee
Using Eq.~(\ref{replica}), one finds for the corresponding contribution to 
the 
correlation function:
\be
S(\omega)|_{p_1=p_2=0} =  
\pi^2+ \sum_{\bf q} {1 \over (D {\bf q}^2 -i \omega)^2} \ .
                                                          \label{u-S1}
\ee
This is the well known perturbative contribution \cite{Altshuler86}. 
The replica non--symmetric manifold with $p_1=p_2=1$ gives:
\be
\langle Z^{(n_1,n_2)}(\hat E)\rangle|_{p_1=p_2=1} = 
n_1 n_2\, { e^{2 \pi i \omega} \over 4 \pi^2 \omega^2}\, 
\prod_{{\bf q}\neq 0} 
\(( {(D {\bf q}^2)^2 \over (D {\bf q}^2)^2 + \omega^2 } \)) \, ,
                                                         \label{U-Z2}
\ee
Differentiating over $\eps_1$ and $\eps_2$ according to Eq.~(\ref{replica}) 
and keeping only the leading contribution 
in $\omega/\Delta \gg 1$, one obtains for the 
corresponding contribution to the correlation function 
\be
S(\omega)|_{p_1=p_2=1}= 
{e^{2 \pi i \omega} \over \omega^2}\,  {\cal D}(\omega)\, , 
                                                        \label{u-S2}
\ee
where we have introduced the spectral determinant
of the diffusion operator, ${\cal D}(\omega)$,  defined as
\be
{\cal D}(\omega) \equiv 
\prod_{{\bf q} \ne 0} \[[1+ \(( {\omega \over D {\bf q}^2}\))^2 \]]^{-1} \ .
                                                \label{D}
\ee
Finally using Eq.~(\ref{S2}) one finds 
\be 
R(\omega) = 
{1\over 2\pi^2} \Re \sum_{\bf q} {1 \over (D {\bf q}^2 -i \omega)^2} +
\frac{\cos(2\pi\omega)}{2\pi^2\omega^2}\, {\cal D}(\omega) \, ,
                                                       \label{u-R}
\ee 
in agreement with Refs.~\cite{Andreev95,Bogomolny96}.

\subsection{Orthogonal ensemble} 
\label{s32}

If the time reversal symmetry is not broken the 
Hubbard--Stratonovich matrix $\hat Q$ appears to be a self--dual real-quaternion 
matrix \cite{Efetov81}. This means that each element $Q_{ij}$ may be written as 
\be
Q_{ij} = \sum\limits_{a=0}^3 Q_{ij}^{a}\tau_a\, 
                                               \label{quater}
\ee
with real $Q_{ij}^{a}$, where 
\be
\tau_0 = \left( \begin{array}{cc}
1 & 0 \\ 0 & 1 \end{array} \right)\, ; \,\,\,  
\tau_1 = \left( \begin{array}{cc}
0 & -i \\ -i & 0 \end{array} \right)\, ; \,\,\,  
\tau_2 = \left( \begin{array}{cc}
0 & -1 \\ 1 & 0 \end{array} \right)\, ; \,\,\,  
\tau_3 = \left( \begin{array}{cc}
-i & 0 \\ 0 & i \end{array} \right)\, .
                                                           \label{tau}
\ee
Moreover $Q_{ji} = (Q_{ij})^\dagger$, where conjugation operation acts on the 
$\tau$ matrices. Such matrices may be diagolized by rotations from the 
{\em symplectic} group \cite{Mehta}, $G=Sp(n_1+n_2)$, 
which is the relevant symmetry group for the GOE.  
Diagonal elements of the $\hat Q$--matrix, $Q_{ii}$ are 
characterized by a single real number, $Q_{ij}^0$, whereas off-diagonal ones 
$Q_{i<j}$ are parameterized by four numbers, $Q_{ij}^{a}$, $a=0,\ldots, 3$. 
Altogether there are $2(n_1+n_2)^2 - (n_1+n_2)$ degrees of freedom 
for each spatial mode, ${\bf q}$.
The measure of the functional integral in Eq.~(\ref{Zorig})  
is given by 
\be
d[\hat Q] = \prod\limits_{\bf q} \[[ \prod\limits_{j} dQ_{jj}^0({\bf q}) 
\prod\limits_{i<j} \prod\limits_{a=0}^3 d Q_{ij}^a({\bf q}) \]] \, .
                                                  \label{0-meas}
\ee 
One easily finds that the number of massive, soft and 
zero modes is 
\bea  
{\cal N}_m &=&
2p_1^2- p_1 + 2(n_1-p_1)^2 - (n_1-p_1) + 2p_2^2 - p_2 + 2(n_2-p_2)^2- (n_2-p_2)
+4 p_1 (n_2-p_2)+4 p_2 (n_1-p_1) \\ 
&=& 2n_1^2+2n_2^2-(n_1+n_2) + 4(p_1-p_2)(p_1-p_2-n_1+n_2) \, ; \nonumber \\
{\cal N}_{s+} &=&
4 p_1 p_2  \, ;\nonumber \\
{\cal N}_{s-} &=&
4 (n_1-p_1) (n_2-p_2) \, ;\nonumber  \\
{\cal N}_z &=& 
4 p_1(n_1-p_1) +4 p_2 (n_2-p_2) \, ,
                                                 \label{o-modes}
\eea
which add up to $2(n_1+n_2)^2- (n_1+n_2)$.
The number of zero modes, ${\cal N}_z$, 
coincides with the number of dimensions 
of the degenerate coset space manifold 
$
{Sp(n_1)}/({Sp(p_1) Sp(n_1-p_1) }) \times 
{Sp(n_2)}/( {Sp(p_2) Sp(n_2-p_2) }) \, .
  $
The volume of this  coset space is calculated in the appendix and is given by 
\be
{\cal V}^{(p_1 p_2)}_{(n_1 n_2)} = 
(4\pi)^{2p_1(n_1-p_1)+2p_2(n_2-p_2)} 
\ E_{n_1}^{p_1} \ E_{n_2}^{p_2} \, ,
                                                          \label{o-volume}
\ee
where 
\be
 E_n^p \equiv {\Gamma(1+n) \over \Gamma(1+p) \Gamma(1+n-p)} 
 \prod\limits_{j=1}^p { \Gamma(1+2j)  \over \Gamma(1+2(n-j+1))} \, .
                                                           \label{o-F}
\ee
Since $E_n^{p>n} = 0$ the sums over $p_1$ and $p_2$ in Eq.~(\ref{Ztot}) may be 
extended up to infinity. The resulting expression may be then continued 
analytically to $n_{1,2} \to 0$ (cf. Ref.  \cite{KM1}).  
In the limit $n\to 0$ the  $E_n^{p}$ symbol, Eq.~(\ref{o-F}), 
is given by 
\be 
E_{n\to 0}^{0} = 1\,; \,\,\,\, E_{n\to 0}^{1} = 2n\,; \,\,\,\, 
E_{n\to 0}^{p\geq 2} = O(n^p)\, .
                                                          \label{o_Fcont}
\ee
Therefore only the terms with $p_{1,2} = 0,1$ may contribute to the correlation 
function $S(\omega)$, Eq.~(\ref{replica}). 
The number of massive modes, in the limit where
$n_{1,2} \to 0$ at fixed $p_1,p_2$, is ${\cal N}_m \to 4 (p_1-p_2)^2$. 
Therefore
the terms with 
$p_1\ne p_2$ may be neglected to leading order  in the parameter $\Delta\tau \ll 
1$.
Like in the unitary case only two terms with $p_1=p_2=0$
and $p_1=p_2=1$  contribute to the generating function. 

The replica symmetric contribution $p_1=p_2=0$ is very similar to the one of 
the unitary ensemble. 
Using Eqs.~(\ref{Ztot}), (\ref{o-modes}) and (\ref{o_Fcont}) one 
finds:
\be
\langle Z^{(n_1,n_2)}(\hat E)\rangle|_{p_1=p_2=0} = 
e^{\pi i(n_2 \eps_2-n_1 \eps_1)} 
\prod\limits_{\bf q} \(( {1 \over D {\bf q}^2-i \omega}\))^{2n_1n_2}\, .
                                                         \label{o-Z1}
\ee
Using Eq.~(\ref{replica}), one finds for the corresponding contribution to 
the correlation function:
\be
S(\omega)|_{p_1=p_2=0} =  
\pi^2+ 2\sum_{\bf q} {1 \over (D {\bf q}^2 -i \omega)^2} \ ,
                                                          \label{o-S1}
\ee
in agreement with the known perturbative calculations \cite{Altshuler86}. 
The replica non--symmetric saddle point manifold, $p_1=p_2=1$, contribute as
\be
\langle Z^{(n_1,n_2)}(\hat E)\rangle|_{p_1=p_2=1} = 
n_1 n_2\, { e^{2 \pi i \omega} \over 4 \pi^4 \omega^4}\, 
\prod_{{\bf q}\ne 0} 
\(( {(D {\bf q}^2)^2 \over (D {\bf q}^2)^2 + \omega^2 } \))^2 \, ,
                                                         \label{o-Z2}
\ee
Differentiating over $\eps_1$ and $\eps_2$ according to Eq.~(\ref{replica}) 
and keeping only the leading contribution 
in $\omega/\Delta \gg 1$, one obtains for the 
corresponding term in the correlation function 
\be
S(\omega)|_{p_1=p_2=1}= 
{e^{2 \pi i \omega} \over \pi^2 \omega^4}\,  {\cal D}^2(\omega)\, , 
                                                        \label{o-S2}
\ee
where the spectral determinant, ${\cal D}(\omega)$,  is defined by 
Eq.~(\ref{D}).
Finally, from Eq.~(\ref{S2}) one finds 
\be 
R(\omega) = 
{1\over \pi^2} \Re \sum_{\bf q} {1 \over (D {\bf q}^2 -i \omega)^2} +
\frac{\cos(2\pi\omega)}{2\pi^4\omega^4}\, {\cal D}^2(\omega) \, ,
                                                       \label{o-R}
\ee 
again in agreement with Refs.~\cite{Andreev95,Bogomolny96}.

\subsection{Symplectic ensemble} 
\label{s33}

If the spin of electrons is taken into account and the strong spin--orbit 
scattering 
is assumed the Hamiltonian of the system acquires a quaternion (symplectic) 
structure 
\cite{Efetov81}. The corresponding symmetry of the $\hat Q$--matrix is the 
{\em orthogonal} one, $G=O(n_1+n_2)$. The $\hat Q$ is a real symmetric matrix 
and 
the integration measure in Eq.~(\ref{Zorig})  is  
\be
d[\hat Q] = 
\prod\limits_{\bf q} \[[ \prod\limits_{i\leq j}  d Q_{ij}({\bf q}) \]] \, .
                                                  \label{s-meas}
\ee 
There are $[(n_1+n_2)^2 + (n_1+n_2)]/2$ real degrees of freedom 
for each spatial mode, ${\bf q}$. The 
number of massive, soft and zero modes is 
\bea  
{\cal N}_m &=&
{1\over 2}[p_1^2+ p_1] + {1\over 2}[(n_1-p_1)^2 + (n_1-p_1)] + 
{1\over 2}[p_2^2 + p_2] + {1\over 2}[(n_2-p_2)^2 + (n_2-p_2)] 
+ p_1 (n_2-p_2)+ p_2 (n_1-p_1) \\ 
&=& {1\over 2}[n_1^2+n_2^2 + n_1+n_2] + (p_1-p_2)(p_1-p_2-n_1+n_2) \, ; 
\nonumber \\
{\cal N}_{s+} &=& 
 p_1 p_2  \, ;\nonumber \\
{\cal N}_{s-} &=& 
(n_1-p_1) (n_2-p_2) \, ;\nonumber  \\
{\cal N}_z &=& 
p_1(n_1-p_1) + p_2 (n_2-p_2) \, ,
                                                 \label{s-modes}
\eea
which correctly add up to $[(n_1+n_2)^2 + (n_1+n_2)]/2$.
The number of zero modes, ${\cal N}_z$, 
coincides with the number of dimensions 
of the degenerate coset space manifold 
$
{O(n_1)}/({O(p_1) O(n_1-p_1) }) \times 
{O(n_2)}/({O(p_2) O(n_2-p_2) })  \, .
$
The volume of this  coset space is calculated in the appendix and is given by 
\be
{\cal V}^{(p_1 p_2)}_{(n_1 n_2)} = 
(2\sqrt{\pi})^{p_1(n_1-p_1)+p_2(n_2-p_2)} 
\ G_{n_1}^{p_1} \ G_{n_2}^{p_2} \, ,
                                                          \label{s-volume}
\ee
where 
\be
 G_n^p \equiv {\Gamma(1+n) \over \Gamma(1+p) \Gamma(1+n-p)} 
 \prod\limits_{j=1}^p { \Gamma(1+j/2)  \over \Gamma(1+(n-j+1)/2)} \, .
                                                           \label{s-F}
\ee
Since $G_n^{p>n} = 0$ the sums over $p_1$ and $p_2$ in Eq.~(\ref{Ztot}) may be 
extended up to infinity. The resulting expression may be continued 
analytically to $n_{1,2} \to 0$ (cf. Ref.  \cite{KM1}).  
The   $G_n^{p}$ symbol, Eq.~(\ref{s-F}), in the limit $n \to 0$ is  
\be 
G_{n\to 0}^{0} = 1\,; \,\,\,\, 
G_{n\to 0}^{1} = \frac{\sqrt{\pi}}{2}n\,; \,\,\,\, 
G_{n\to 0}^{2} = - \frac{1}{4}n\,; \,\,\,\, 
G_{n\to 0}^{p\geq 3} = O(n^{[(p+1)/2]})\, ,
                                                          \label{s_Fcont}
\ee
where $[x]$ denotes integer part of $x$. 
Therefore only the terms with $p_{1,2} = 0,1,2$ contribute to the correlation 
function $S(\omega)$, Eq.~(\ref{replica}). 
The number of massive modes, in the limit where
$n_{1,2} \to 0$ at fixed $p_1,p_2$, is ${\cal N}_m \to (p_1-p_2)^2$, making 
terms with $p_1\ne p_2$ small in the parameter $\Delta\tau \ll 1$.
One therefore finds three  relevant contributions to the generating function: 
$p_1=p_2=0$, $p_1=p_2=1$ and $p_1=p_2=2$. 

The replica symmetric contribution $p_1=p_2=0$ comes almost without changes.
Employing Eqs.~(\ref{Ztot}), (\ref{s-modes}) and (\ref{s_Fcont}) one 
finds:
\be
\langle Z^{(n_1,n_2)}(\hat E)\rangle|_{p_1=p_2=0} = 
e^{\pi i(n_2 \eps_2-n_1 \eps_1)} 
\prod\limits_{\bf q} \(( {1 \over D {\bf q}^2-i \omega}\))^{n_1n_2/2}\, .
                                                         \label{s-Z1}
\ee
  From Eq.~(\ref{replica}), one finds for the corresponding contribution to 
the correlation function:
\be
S(\omega)|_{p_1=p_2=0} =  
\pi^2+ {1\over 2}\sum_{\bf q} {1 \over (D {\bf q}^2 -i \omega)^2} \ ,
                                                          \label{s-S1}
\ee
in agreement with  Ref. \cite{Altshuler86}. 
The first replica non--symmetric manifold, $p_1=p_2=1$, results in 
\be
\langle Z^{(n_1,n_2)}(\hat E)\rangle|_{p_1=p_2=1} = 
n_1 n_2\, { e^{2 \pi i \omega} \over 8 \omega}\, 
\prod_{{\bf q}\ne 0} 
\(( {(D {\bf q}^2)^2 \over (D {\bf q}^2)^2 + \omega^2 } \))^{1/2} \, ,
                                                         \label{s-Z2}
\ee
Differentiating over $\eps_1$ and $\eps_2$ according to Eq.~(\ref{replica}) 
and keeping only the leading contribution 
in $\omega/\Delta \gg 1$, one obtains for the 
corresponding contribution to the correlation function 
\be
S(\omega)|_{p_1=p_2=1} = 
{\pi^2 e^{2 \pi i \omega} \over 2 \omega}\,  
\sqrt{ {\cal D}(\omega) }\, , 
                                                        \label{s-S2}
\ee
where the spectral determinant, ${\cal D}(\omega)$,  is defined by 
Eq.~(\ref{D}).
Finally, the second replica non--symmetric manifold, $p_1=p_2=2$, gives 
\be
\langle Z^{(n_1,n_2)}(\hat E)\rangle|_{p_1=p_2=2} = 
n_1 n_2\, { e^{4 \pi i \omega} \over (4\pi)^4 \omega^4}\, 
\prod_{{\bf q}\ne 0} 
\(( {(D {\bf q}^2)^2 \over (D {\bf q}^2)^2 + \omega^2 } \))^2 \, ,
                                                         \label{s-Z3}
\ee
and consequently 
\be
S(\omega)|_{p_1=p_2=2} = 
{e^{4 \pi i \omega} \over 16 \pi^2 \omega^4}\,  {\cal D}^2(\omega)\, . 
                                                        \label{s-S3}
\ee
Using Eq.~(\ref{S2}) one finds 
\be 
R(\omega) = 
{1\over 4 \pi^2} \Re \sum_{\bf q} {1 \over (D {\bf q}^2 -i \omega)^2} +
\frac{\cos(2\pi\omega)}{4 \omega}\, \sqrt{ {\cal D}(\omega)} + 
\frac{\cos(4\pi\omega)}{32\pi^4 \omega^4}\, {\cal D}^2(\omega) \, ,
                                                       \label{s-R}
\ee 
again in agreement with Refs.~\cite{Andreev95,Bogomolny96}.

\section{Discussion of the Results}
\label{s4}

Let us briefly discuss the  energy scales, the
approximations involved in the calculations 
and their range of validity. There are four important energy scales: the mean 
level 
spacing, $\Delta$; the Thouless energy, $E_c=\hbar D/L^2$ ($L$ is the system 
size);
the inverse scattering time, $\hbar/\tau$; and the chemical potential, $\mu$. In 
the 
calculations above, the following hierarchy was assumed: 
$\Delta\ll E_c\ll \hbar/\tau\ll \mu$. The condition  $\hbar/\tau\ll \mu$, 
which  means 
that the disorder is weak enough,  was used to evaluate momentum sums by contour 
integration. The inequality  $E_c\ll \hbar/\tau$, which  is equivalent to $L$
much larger than the mean free path $l$, 
 tells that the system is in the diffusive regime. It 
was 
used to derive the diffusive dispersion law in Eq.~(\ref{sums}). Finally, 
$g=E_c/\Delta$ is the dimensionless conductance, and the condition that $g\gg 1$ 
means 
that 
the system is metallic. This condition was used to calculate integrals over zero 
modes with 
${\bf q}\ne 0$ in the saddle point approximation. One more inequality was 
assumed in our derivation, the fact that the difference in energies
$\omega$ is much larger than the level spacing $\Delta$.
 This is a technical assumption, which allowed us to evaluate 
soft modes integrals by the saddle point technique. It would be interesting 
 to perform the calculations without  this last assumption, extending thus
 the results to arbitrarily small $\omega$. 
  
Our calculations give the correlation  as functions of $\omega$ in 
the form  of a finite sum of oscillating harmonics (two in the 
GOE 
and GUE and three in the GSE), with  $\omega$ dependent amplitudes. 
The 
set of harmonics is exact and has to do only with the symmetry of the problem,
specifically with the volumes of the relevant coset spaces. The amplitudes, on 
another hand, were obtained in the saddle point approximation only. Using 
our formulation, one may develop a perturbation theory near the 
replica non-symmetric saddle points, much in the same way as it  was done near 
the replica symmetric one, see e.g. Ref. \cite{Lerner98}. From such a 
perturbation theory one may obtain a systematic expansion of 
the amplitudes of the oscillatory terms, in powers of 
$\Delta/\omega < 1$. 

We would like to point out  striking similarities between our replica approach 
and 
the SUSY one of Ref. \cite{Andreev95} which was also based on the saddle point 
calculations. In particular the list of modes is the same. 
In the SUSY case, the zero modes and soft modes are respectively
 associated with the 
rotations inside the fermionic block and between fermions and bosons.
In some sense our $p$ and $n-p$ replica blocks are similar to the  bosonic and 
fermionic 
blocks of the SUSY theory. To appreciate better this analogy, one would need a 
more 
detailed understanding of the mathematical structure of the theory. In 
particularly, 
one would like to define the unitary (or other) group, $U(n)$, for non--integer 
$n$ and trace its relation to the graded symmetry. Another interesting problem
is to appreciate better connections to the semi-classical method of 
Ref. \cite{Bogomolny96}.   

The existence of the  replica non-symmetric saddle points
opens two very important questions. One concerns their relevance to
the renormalization group treatment of the localization problem
for one electron. Another, 
even more challenging one, is to extend the replica theory of interacting 
electrons \cite{Finkelstein83} to account for new saddle points.

\section*{Acknowledgments}
\label{ackno}

We benefited a lot from  numerous discussions with A. Andreev, 
I. Gruzberg and S. Nishigaki. 
We are grateful to I. Yurkevich and I. Lerner for
 communicating to us
their results
prior to publication and M. Zirnbauer for useful correspondence. 
This research was supported in part by the
National Science Foundation under grant No. PHY94-07194. A.K. was supported by 
NSF grant  DMR 93--0801.

 \section*{Appendix: Zero modes and volume of the coset space}

 \subsection{Unitary case}
The  manifold ${\cal M}_{(p_1p_2)}$  of saddle point matrices $\hat Q$ is 
generated
by unitary transformations $U$ of $U(n_1) \times U(n_2)$,  
applied to the diagonal matrix 
$\hat \Lambda_{(p_1p_2)}$, cf. Eq.~(\ref{Usp}). 
We must first find which choices of $V_1,V_2$ actually
change the $\hat Q$ matrix. A general unitary transformation $V$
of $U(n_1)$ can be written as a product $ R W$ where the matrix $W$ has the 
block
diagonal structure: 
\be
W= \left( 
\begin{array}{cc}
W_{p_1}  & 0 \\
0        &   W_{n_1-p_1}
\end{array} 
\right)
\ee
where $W_{p_1}$ and $W_{n_1-p_1}$ are unitary matrices of size $p_1$ and 
$n_1-p_1$ 
respectively. 
The matrix $W$ belongs to the subgroup $U(p_1) \times U(n_1-p_1)$ of $U(n_1)$ 
which 
leaves
the saddle point matrix invariant. 
The `proper' $V_1$ transformations which change
the matrix $\hat Q$  while staying on the saddle point manifolds are thus the
elements $R_1$ of the coset space $\frac{U(n_1)}{U(p_1) U(n_1-p_1)}$, and 
similarly
the proper $V_2$ transformations are elements $R_2$ 
which belong to $\frac{U(n_2)}{U(p_2) U(n_2-p_2)}$.
 
To compute the volume of the set of proper $R$ transformations in $U(n)$
(here $n$ stands for either $n_1$ or $n_2$), 
we start from the usual decomposition \cite{Mehta} of the integral
 over the group of $n \times n$ Hermitian matrixes $X$ in terms of
 the $n$ eigenvalues  $x_j$, and the unitary
 transformation $V$ such that $X=V^{-1} (\mbox{diag} \{ x_1,...x_n \}) V$:
\be
 d\rho_n (X)= \prod_{j=1}^n d x_j \prod_{j=1}^{n-1} \theta( x_{j+1}
 -x_j) \prod_{1 \le j <k \le n} (x_j-x_k)^2  \ \ d\rho_n (V) \ .
 \label{decomp_meas}
 \ee
In this integral we have ordered the eigenvalues (the $\theta$ function is 
Heavyside's
step function), in such a way that the integral over $V$ scans the whole 
set of allowed
unitary transformations. We can compute the normalization of the `angular' 
measure
for instance by integrating a Gaussian function:
\be
I\equiv \int d \rho_n (X) \exp\((-{1 \over 2} \Tr X^2\)) = \pi^{n^2/2} 
2^{n/2} 
\ee
which can be computed using the above decomposition and the Selberg's integral 
\cite{Mehta}:
\be
I={ 1 \over n!}
\[[\int d\rho_n (V) \]]
\ \int \prod_{j=1}^n d x_j  \prod_{j<k}  (x_j-x_k)^2 \exp\((-{1 \over 2}
 \sum_j x_j^2\))\\
 = { 1 \over n!} \[[\int d\rho_n (V) \]]
  (2 \pi)^{n/2} \prod_{j=1}^n \Gamma(j+1) \ .
  \label{volume_U}
 \ee 
 Therefore one gets the normalization of the integral over the angular
 measure:
 \be
{\cal V}^U_n \equiv \int d\rho_n (V)= \pi^{(n^2-n)/2} { n! \over 
\prod_{j=1}^n \Gamma(j+1)} \ .
 \label{Uvol}
 \ee
 This result is easily checked by a direct counting argument: the
 choice of $V$ is a choice of a Hermitian basis. The first vector of the basis 
 is an arbitrary unit vector, the corresponding volume of integration is thus 
 $S_{2 n}/2 \pi$ where $S_d= d \pi^{d/2} / \Gamma(1+d/2)$ is the volume of the
 $d$ dimensional unit sphere, and the division by $2 \pi$ deals with a 
 global phase choice. The second unit vector of the basis must be orthogonal
 to the  first one, which fixes two real conditions, and its volume is thus 
 $S_{2 n -2}/2 \pi$.
 After iterating this construction, one gets the result (\ref{Uvol}).
 
 We now decompose $V=R W$, and  the angular integral $d \rho_n (V)$ as:
 \be
 d\rho_n (V)= d\rho_p (W_p) d\rho_{n-p} (W_{n-p}) d\rho_{n,p} (R)
                                                  \label{decomp}
 \ee
 This defines the measure $d\rho_{n,p}$ in the $2p(n-p)$ space of the 
  proper transformations $R$. The normalization of this measure  is:
 \be
 \int 
 d\rho_{n,p} (R) = {{\cal V}^U_n \over {\cal V}^U_p {\cal V}^U_{n-p}}
 =\pi^{p(n-p)} \ F_n^p
                                                  \label{decomp_ang}
 \ee
 where we have introduced the 
 symbol $F_n^p$ defined by:
 \be
 F_n^p \equiv {n! \over p! (n-p)!} { \prod_{j=1}^p \Gamma(j+1) \ \ 
 \prod_{j=1}^{n-p}\Gamma(j+1) \over \prod_{j=1}^n \Gamma(j+1)} \ .
                                                    \label{aF}
 \ee

 We can now go on to the exact evaluation of the zero mode integrals.
 We keep within the subspace of the space independent $\hat Q$ matrices  
(${\bf q}=0$  modes)
which are the only modes having a zero eigenvalue sector. Clearly the
zero modes integrals factorize into two independent pieces, associated with each
of the two coset spaces $\frac{U(n_1)}{U(p_1) U(n_1-p_1)}$ and  
$\frac{U(n_2)}{U(p_2) 
U(n_2-p_2)}$.
We can compute each such piece by working with a $n \times n$ Hermitian
matrix $X$ and computing:
\be
Z^{(n)} \equiv \int d\rho_n(X) \exp\((-{\pi \nu \over  4 \tau} \Tr X^2
+ \Tr \ln \(( E  + {i \over 2 \tau }X\)) \))
\ee
We expand around the saddle point manifold generated by
 $X= \Lambda_p\equiv \mbox{diag} \{\lambda_1...\lambda_n \}=\mbox{diag} 
\{-1...,-1,1,...1 \}$
by writing:
\be
X=R \Lambda_p  R^{-1} +\delta X\ \ , \ \ 
\delta X=RW (\mbox{diag}\{x_1,...,x_n\}) W^{-1} R^{-1} \ ,
\ee
where $W$, as above, is in $U(p) \times U(n-p)$ and $R$ is a proper 
transformation. 
Using the decompositions of the measure defined in 
Eqs.~(\ref{decomp_meas}), (\ref{decomp}),
one obtains 
\bea
 Z^{(n)} &=& \int dx_1...dx_p \prod_{1 \le i < j \le p} (x_i-x_j)^2 
\  \prod_{j=1}^{p-1} \theta (x_{j+1}-x_j) \ \int d \rho_p(V_p) \times \\ \nn
&&\int dx_{p+1}...dx_n \prod_{p+1 \le i < j \le n} (x_i-x_j)^2 
\ \prod_{j=p+1}^{n-1} \theta (x_{j+1}-x_j) \ \int d \rho_{n-p}(V_{n-p})\times 
                                                                \\ \nn
&& \int d\rho_{n,p}(R) \prod_{i=1}^p \prod_{j=p+1}^n (-2+x_i-x_j)^2 
\ \theta(2+x_{p+1}-x_p) \times
                            \\ 
&& \exp \left\{  -A_p+{\pi \nu \over  4 \tau} \sum_{j=1}^n x_j^2
+{1 \over 8 \tau^2} \Tr \(( (E+i \Lambda_p/2\tau)^{-1} \delta X 
(E+i \Lambda_p/2\tau)^{-1} \delta X \)) \right\} \ ,
                                                            \label{big}
\eea
where $A_p$ is the saddle point action:
\be
A_p=
{\pi \nu \over  4 \tau} \Tr \Lambda_p^2 + 
\Tr \ln \(( E  + {i }\Lambda_p/2 \tau \))\ .
                                                           \label{biga}
\ee
The integral in Eq.~(\ref{big}) can be simplified by the following observations:
the integrals over $x_i$ are all massive modes, and thus one can assume
that $|x_i-x_j| \ll 1$. Therefore the third line of Eq.~(\ref{big}) is just a 
constant, equal to $\pi^{p(n-p)} F_n^p 2^{2p(n-p)}$. Apart from this constant, 
the
rest of Eq.~(\ref{big}) is nothing but the integrals over the massive modes.

What we have shown here is that, in the sector ${\bf q}=0$ 
of uniform fluctuations,
the exact integral over the saddle point manifold (the zero mode directions)
 gives a factor:
\be
 (4\pi)^{p_1(n_1-p_1)+p_2(n_2-p_2)} \ F_{n_1}^{p_1}
  \ F_{n_2}^{p_2} 
\ee

\subsection{Orthogonal case}

We shall not repeat here all the steps of the previous computation, they run
in exactly the same way. We just give the main modifications.
 The integral over the symplectic group, generalizing 
Eq.~(\ref{volume_U}) is equal to:
 \be
 {\cal V}^S_n=\int d\rho_n(V)= n! \((\int d \rho_n(X) \exp\((-{X^2 \over 
2}\))\))
 \(( \int \prod\limits_{j=1}^n 
dx_j \prod_{j<k} (x_j-x_k)^4 \exp\[[-{1 \over 2} \sum_j x_j^2\]]  \))^{-1}
 \ee
 and the computation of Selberg's integral gives the volume:
 \be
 {\cal V}^S_n= n!\, {\pi^{2n^2-3n/2} 2^{n} \over \prod_{j=1}^n \Gamma(1+2j)} \ .
 \ee
 The ratio of volumes is
 \be
 {{\cal V}^S_n \over {\cal V}^S_p {\cal V}^S_{n-p}}= \pi^{2p(n-p)} E_n^p
 \ee
 where
 \be
  E_n^p \equiv {n! \over p! (n-p)!} { \prod_{j=1}^p \Gamma(2j+1) \ \ 
 \prod_{j=1}^{n-p}\Gamma(2j+1) \over \prod_{j=1}^n \Gamma(2j+1)} \, . 
 \ee
 Finally the factor coupling the eigenvalues  $j\le p$  to those $j>p$ in
 the analog of Eq.~(\ref{big}) becomes:
 \be
 \prod_{i=1}^p \prod_{j=p+1}^n (-2+x_i-x_j)^4
 \ee
 so that the final integral over the zero mode manifold is:
 \be
 (4\pi)^{2[ p_1(n_1-p_1)+p_2(n_2-p_2)]} E_{n_1}^{p_1}
 E_{n_2}^{p_2} 
 \ee

\subsection{Symplectic case}

 We just give again the main modifications.
 The integral over the orthogonal group, generalizing 
Eq.~(\ref{volume_U}) is equal to:
 \be
 {\cal V}^O_n=\int d\rho_n(V)= n! \((\int d \rho_n(X) \exp\((-{X^2 \over 
2}\))\))
 \(( \int \prod_{j=1}^n dx_j \prod_{j<k} |x_j-x_k| 
\exp\[[-{1 \over 2} \sum_j x_j^2\]]  \))^{-1}
 \ee
 where $X$ is a real symmetric matrix.
 The computation of Selberg's integral gives the volume:
 \be
 {\cal V}^O_n= n! {\pi^{n^2/4+n/4)} 2^{n^2/4-5n/4} \over \prod_{j=1}^n 
 \Gamma(1+j/2)} \ .
 \ee
 The ratio of volumes is
 \be
 {{\cal V}^O_n \over {\cal V}^O_p {\cal V}^O_{n-p}}= (2\pi)^{p(n-p)/2} G_n^p
 \ee
 where
 \be
  G_n^p \equiv {n! \over p! (n-p)!} { \prod_{j=1}^p \Gamma(j/2+1) \ \ 
 \prod_{j=1}^{n-p}\Gamma(j/2+1) \over \prod_{j=1}^n \Gamma(j/2+1)} \, . 
 \ee 
 Finally the factor coupling the eigenvalues  $j\le p$  to those $j>p$ in
 the analog of Eq.~(\ref{big}) becomes:
 \be
 \prod_{i=1}^p \prod_{j=p+1}^n |-2+x_i-x_j|
 \ee
 so that the final integral over the zero mode manifold is:
 \be
 (4\pi)^{[ p_1(n_1-p_1)+p_2(n_2-p_2)]/2} \ G_{n_1}^{p_1}\ 
 G_{n_2}^{p_2} 
 \ee

%\ecols
\end{document}